\documentclass[conference]{IEEEtran}
\IEEEoverridecommandlockouts
\usepackage{cite}
\usepackage{hyperref}
\usepackage{amsmath,amssymb,amsfonts}
\usepackage{algorithmic}
\usepackage{graphicx}
\usepackage{textcomp}
\usepackage{xcolor}
\usepackage{stfloats}
\def\BibTeX{{\rm B\kern-.05em{\sc i\kern-.025em b}\kern-.08em
    T\kern-.1667em\lower.7ex\hbox{E}\kern-.125emX}}
\begin{document}

\title{Managed Network Services for Exascale Data Movement Across Large Global Scientific Collaborations}

\author{\IEEEauthorblockN{Frank W\"urthwein}
\IEEEauthorblockA{
\textit{Department of Physics} \\
\textit{UC San Diego}\\
La Jolla, USA \\
fkw@ucsd.edu}
\and
\IEEEauthorblockN{Jonathan Guiang}
\IEEEauthorblockA{
\textit{Department of Physics} \\
\textit{UC San Diego}\\
La Jolla, USA \\
jguiang@ucsd.edu}
\and
\IEEEauthorblockN{Aashay Arora}
\IEEEauthorblockA{
\textit{Department of Physics} \\
\textit{UC San Diego}\\
La Jolla, USA \\
aaarora@ucsd.edu}
\and
\IEEEauthorblockN{Diego Davila}
\IEEEauthorblockA{
\textit{SDSC} \\
\textit{UC San Diego}\\
La Jolla, USA \\
didavila@ucsd.edu}
\and
\IEEEauthorblockN{John Graham}
\IEEEauthorblockA{
\textit{SDSC} \\
\textit{UC San Diego}\\
La Jolla, USA \\
jjgraham@ucsd.edu}
\and
\IEEEauthorblockN{Dima Mishin}
\IEEEauthorblockA{
\textit{SDSC} \\
\textit{UC San Diego}\\
La Jolla, USA \\
dmishin@ucsd.edu}
\and
\IEEEauthorblockN{Thomas Hutton}
\IEEEauthorblockA{
\textit{SDSC} \\
\textit{UC San Diego}\\
La Jolla, USA \\
hutton@sdsc.edu}
\and
\IEEEauthorblockN{Igor Sfiligoi}
\IEEEauthorblockA{
\textit{SDSC} \\
\textit{UC San Diego}\\
La Jolla, USA \\
isfiligoi@sdsc.edu}
\and
\IEEEauthorblockN{Harvey Newman}
\IEEEauthorblockA{
\textit{Department of Physics} \\
\textit{Caltech}\\
Pasadena, USA \\
newman@hep.caltech.edu}
\and
\IEEEauthorblockN{Justas Balcas}
\IEEEauthorblockA{
\textit{Department of Physics} \\
\textit{Caltech}\\
Pasadena, USA \\
jbalcas@caltech.edu}
\and
\IEEEauthorblockN{Tom Lehman}
\IEEEauthorblockA{
\textit{Energy Sciences Network} \\
\textit{Lawrence Berkeley National Laboratory}\\
Berkeley, USA \\
tlehman@es.net}
\and
\IEEEauthorblockN{Xi Yang}
\IEEEauthorblockA{
\textit{Energy Sciences Network} \\
\textit{Lawrence Berkeley National Laboratory}\\
Berkeley, USA \\
xiyang@es.net}
\and
\IEEEauthorblockN{Chin Guok}
\IEEEauthorblockA{
\textit{Energy Sciences Network} \\
\textit{Lawrence Berkeley National Laboratory}\\
Berkeley, USA \\
chin@es.net}
}

\maketitle

\begin{abstract}
Unique scientific instruments designed and operated by large global collaborations are expected to produce Exabyte-scale data volumes per year by 2030. 
These collaborations depend on globally distributed storage and compute to turn raw data into science. 
While all of these infrastructures have batch scheduling capabilities to share compute, Research and Education networks lack those capabilities. 
There is thus uncontrolled competition for bandwidth between and within collaborations. 
As a result, data ``hogs" disk space at processing facilities for much longer than it takes to process, leading to vastly over-provisioned storage infrastructures. 
Integrated co-scheduling of networks as part of high-level managed workflows might reduce these storage needs by more than an order of magnitude. 
This paper describes such a solution, demonstrates its functionality in the context of the Large Hadron Collider (LHC) at CERN, and presents the next-steps towards its use in production.
\end{abstract}


\section{Introduction}

We envision a future where networks are predictable and accountable, both between and within collaborations, and higher level services can reliably express priority between PB-scale flows, while smaller flows continue as today. 
We thus expect that roughly 25\% of the network bandwidth across infrastructures like the Energy Sciences Network (ESnet)~\cite{ESnet} and LHCONE/LHCOPN~\cite{lhcone, lhcopn} remains reserved for ``free-for-all'' operations, while the remainder is available for scheduled traffic if needed. 
Free-for-all can exceed its 25\% on a given network segment when there are no scheduled transfers consuming the remaining 75\%. 

To facilitate this, each end-point storage infrastructure implements a fixed set of IPv6 subnets that can be scheduled in analogy to “batch slots” to consume the total network bandwidth provisioned at the site. 
One such slot is reserved for free-for-all at all times, while the remaining are dynamically attached to end-to-end VPNs between storage sites. 
The mental model is a set of Data Transfer Nodes (DTN), each of which supports all slots, and all of which connect to the same backend filesystem. 
Dynamically allocated VPNs connect a slot at each of two sites across all DTNs the sites operate. 
A single slot can thus be assigned the totality of a site’s network bandwidth capabilities to a single high priority data flow, in principle. 

Higher level collaboration-specific data management systems (DMS) request bandwidth from a singular network scheduling interface (NSI). 
The concept here is that the NSI provides a “promise”  to the DMS of bandwidth between sites for a fixed amount of time to complete the transfer of a fixed volume of data. 
In principle, the NSI could also update promises as demand on the network changes or segments have reduced capacity for other reasons. 
DMS and NSI are thus communicating regularly, changing past promises as needed and possible, both up and down. 
To be able to make such promises, the NSI has access to the known provisioned bandwidth limits of all participating endpoints and network segments, in addition to the complete allocatable network topology, as well as the present commitments of bandwidth via active promises. 
The NSI reviews all active requests and promises on a regular basis, and adjusts promises up or down as allowed given the availability of network bandwidth. 

Implicit is the assumption that flows last much longer than the transient time to dynamically change VPNs. 
It is thus possible to measure achieved throughput over time and reconcile that against the promises made. 
This accountability is an essential conceptual element of our vision as it allows for data analytics to find problems as promises on some routes, involving some network segments, or some sites are systematically not quite met. 
Independent monitoring traffic via systems like perfSonar \cite{Campana_2014} are thus augmented by the accounting of the promises to understand the relevant high level performance characteristics. 

This basic high-level vision is independent of the detailed algorithms used to allocate promises to requests. 
This is motivated, given we are addressing the needs of Exascale data movement, by the fact that moving an exabyte of data takes 100 days even at Tb/s transfer rates. 

\begin{figure*}[t]
\centerline{\includegraphics[width=.75\textwidth]{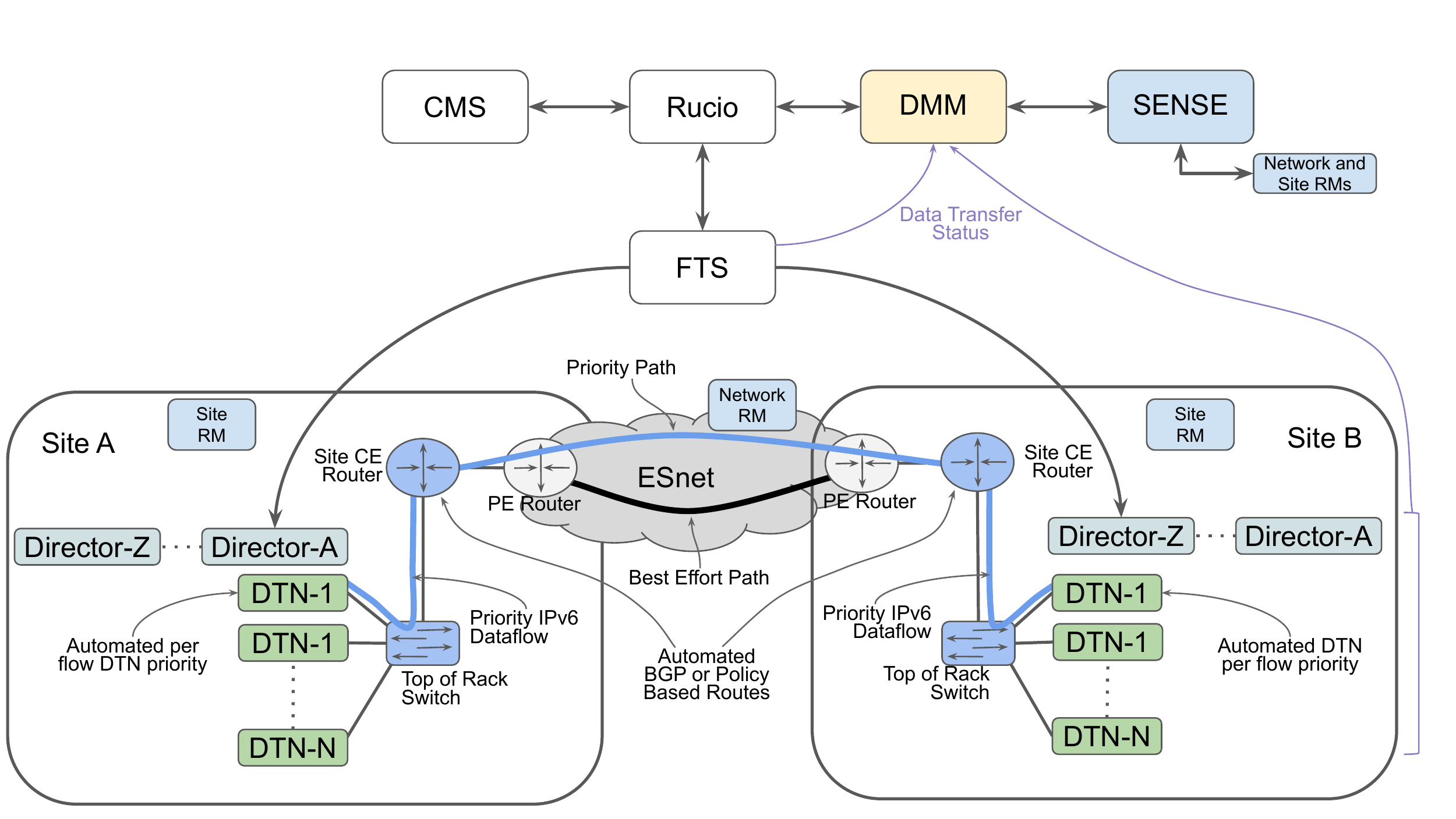}}
\caption{
  Diagram of the Rucio-SENSE interoperation prototype as presented to the 2022 Snowmass conference \cite{snowmass}. 
  A typical workflow for supporting priority data transfers would proceed as follows: 
  An operator at CMS initiates a rule in Rucio; 
  Rucio sends information about this rule, including the priority, to DMM; 
  based on this information, DMM negotiates a bandwidth provision from SENSE; 
  SENSE begins constructing a guaranteed-bandwidth path between the source and destination sites; 
  DMM sends the IPv6 endpoints of that path back to Rucio; 
  Rucio injects the endpoints into the original FTS request; FTS initiates the data transfers. 
  Importantly, when a given workflow ends, DMM can track whether a network promise was fully fulfilled and utilized, compare that to the monitored performance of the sites involved, and thus identify malfunctions in either the northbound or southbound components in the architecture shown here.
}
\label{fig:proto-workflow}
\end{figure*}

\section{Current Status}
\subsection{Deployed Infrastructure}

Our initial scientific target community is the CMS collaboration at the Large Hadron Collider (LHC) at CERN \cite{CERNWeb} because by itself, CMS expects more than half an exabyte of new data for each year of LHC operations during the High-Luminosity LHC era from about 2028-2040 \cite{CMSProjections}. 
Thus, innovative network services have been identified as a necessary improvement towards this era \cite{Albrecht2019, Zurawski2021}. 
As DMS, CMS uses Rucio \cite{rucio2019}, a software framework designed to organize and manage exascale scientific data volumes using customizable policies. 
Rucio is the de-facto standard DMS for the majority of global scientific data-rich collaborations in nuclear, particle and astrophysics. 
Our work targeting CMS is thus potentially relevant to all collaborations using Rucio as their DMS.  

As an NSI, we base ourselves on SENSE \cite{MONGA2020181}, developed by a collaboration between teams at ESnet and Caltech. 
To facilitate rapid prototyping, and minimize modifications of either Rucio or SENSE, we add a layer between Rucio and SENSE that we call the Data Movement Manager (DMM). 
Long term, we expect to work with the developers of Rucio and SENSE to decide where new functionality to implement our vision should be located permanently. 
We thus view DMM as a ``temporary vessel" that we include only in the prototypical architecture presented here. 
In addition, we use XRootD \cite{XRootD} to implement the slot concept at sites.

Figure \ref{fig:proto-workflow} depicts this conceptual architecture. 
FTS \cite{Ayllon_2014} is used to manage the actual data transfer, just like in the production infrastructure used by the LHC experiments and others. 
For our prototype, no changes were necessary in either FTS or XRootD. 
The slot concept at sites could be implemented by an appropriate configuration of the XRootD deployment without any actual changes in the XRootD software. 
Additional technical details may be found in \cite{snowmass, pearc22}.

\begin{figure*}[t]
\centerline{\includegraphics[width=.95\textwidth]{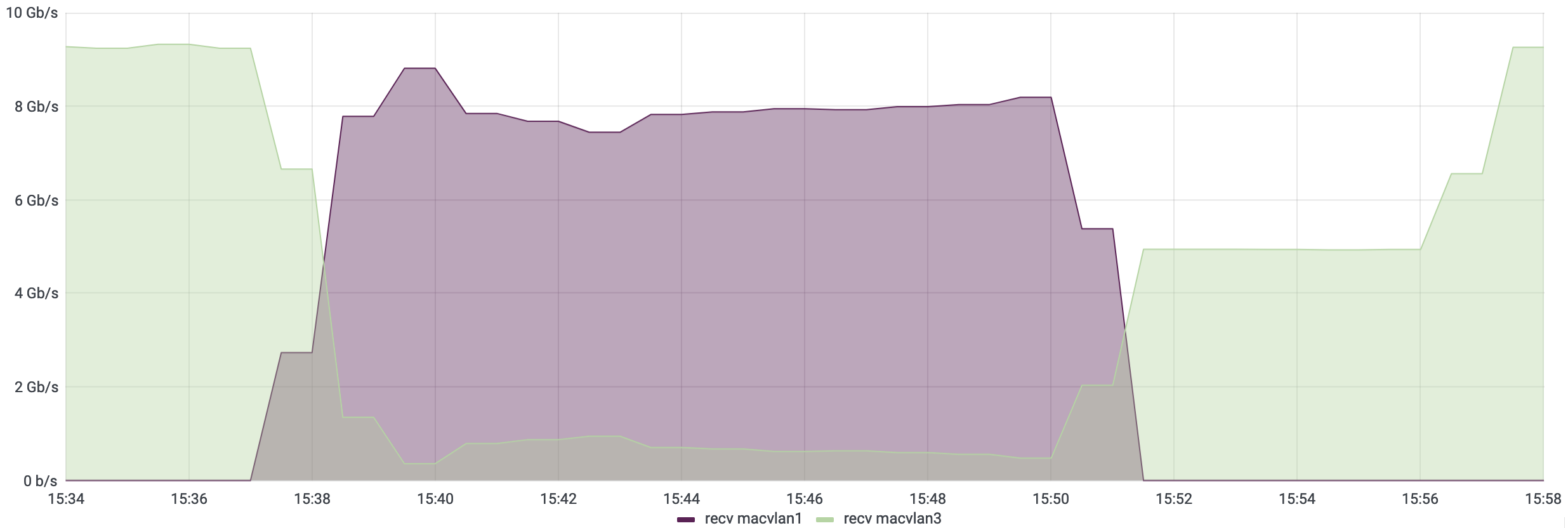}}
\caption{
  Measured throughput of network traffic from UCSD to Caltech is plotted as a function of time in different colors corresponding to the virtual interface on the Caltech DTN. 
  Best-effort traffic flows through the virtual interface plotted in green, while priority flows through the one plotted in purple. 
  To start, best-effort traffic is generated by IPerf, yielding an average throughput of 9.2~Gb/s. 
  At 15:37, Rucio prepares and submits a total of 750 priority file transfers (1~GB each) between UCSD and Caltech, triggering the initialization of a 7~Gb/s SENSE guaranteed-bandwidth service. 
  SENSE finishes implementing the service at 15:38--this in turn restricts best-effort traffic to a maximum of 5~Gb/s.  
  From 15:38 to 15:51, the priority traffic sees at least 7~Gb/s, while best-effort is correctly throttled to a minimum of 100~Mb/s. 
  When the file transfers finish at 15:52, best-effort recovers to the maximum of 5~Gb/s until the SENSE service is terminated at 15:56. 
  Importantly, this 5~Gb/s maximum for best-effort traffic is a tunable parameter and does not represent the setting we plan to use in production.
}
\label{fig:proto-test}
\end{figure*}

Since the publication of the aforementioned work, this prototype has been completely tested from end to end.
For this test, the entire architecture in Figure \ref{fig:proto-workflow} was deployed--though CENIC~\cite{CENIC} was used in place of ESnet--where Site~A was hosted at UC San Diego and Site~B at Caltech, with two DTNs at each site.
To start, best-effort (previously referred to as ``free-for-all'') traffic was initialized and maintained by IPerf~\cite{iperf}.
Next, 750 files (1~GB each) were registered and prepared for transfer by Rucio. 
These transfers were given priority status, triggering a call to DMM that began the construction of a SENSE priority service. 
This service was appropriately routed (Table \ref{tab:proto-route}), and the files were successfully moved across it from UC San Diego to Caltech via the full Rucio-FTS stack. 
We show in Fig. \ref{fig:proto-test} that the best-effort traffic was appropriately throttled, and that the priority traffic was given at least its requested bandwidth of 7 Gb/s. 
Once the transfers were complete, the service was closed by DMM. 
This test proves definitively that the separate components of our prototype work together in concert and that the fundamental action of SENSE has been successfully implemented.

\begin{table}[htbp]
\caption{Traceroute output}
\begin{center}
\begin{tabular}{|c|l|}
\hline
 & \textbf{UCSD-to-Caltech Traceroute$^{\mathrm{a}}$} \\
\hline
        & \scriptsize{\texttt{1  2001:48d0:3001:111::1}} \\
        & \scriptsize{\texttt{2  2001:48d0:fff:990::2}} \\
Before  & \scriptsize{\texttt{3  hpr-lax-hpr--sdsc-10ge.cenic.net}} \\
        & \scriptsize{\texttt{4  hpr--caltech-ul--lax-agg10.cenic.net}} \\
        & \scriptsize{\texttt{5  2605:d9c0:0:ff02::1}} \\
        & \scriptsize{\texttt{6  sense-origin-01.ultralight.org}} \\
\hline
        & \scriptsize{\texttt{1  2001:48d0:3001:111::1}} \\
 After  & \scriptsize{\texttt{2  fc00:3600::17}} \\
        & \scriptsize{\texttt{3  sense-origin-01.ultralight.org}} \\
\hline
\multicolumn{2}{l}{$^{\mathrm{a}}$Showing only the hostname of each hop.}
\end{tabular}
\label{tab:proto-route}
\end{center}
\end{table}

\subsection{Simulation}
We consider developing effective policies on how bandwidth should be shared one of the key conceptual challenges long term. 
We are concerned that ``fair sharing" of network bandwidth is more complex than sharing compute resources in a batch cluster because routes tend to overlap on segments in the network, and end-to-end transfers between sites generally can be accomplished across multiple routes. 
  
To facilitate exploration of this problem space, we have started developing a simulation of the entire system that allows actual instances of Rucio and DMM to be used in the simulation against simulated behaviour of SENSE, the data transfer infrastructure comprised of FTS and XRootD, and the full complexity of the network topology. The goal of this simulation is three-fold. 
First, we want to be able to validate our understanding of what we observe on the testbed against simulation, especially as we add more and more sites to the testbed. 
Second, we want to be able to play back actual annual sequences of Rucio requests and policies with different underlying network allocation policies in SENSE and DMM to demonstrate the benefit of our vision. 
CMS has detailed records of traces for past Rucio requests as well as FTS managed data transfers that we have access to. Performing simulated playbacks, and comparing simulated completions of Rucio requests under different network bandwidth allocation policies is thus in principle possible. 
And third, we want to engage with computer science researchers on developing frameworks of policies for network bandwidth allocation that would be effective in the HL-LHC future if implemented in this system.

\begin{figure}[htbp]
\centerline{\includegraphics[width=.95\linewidth]{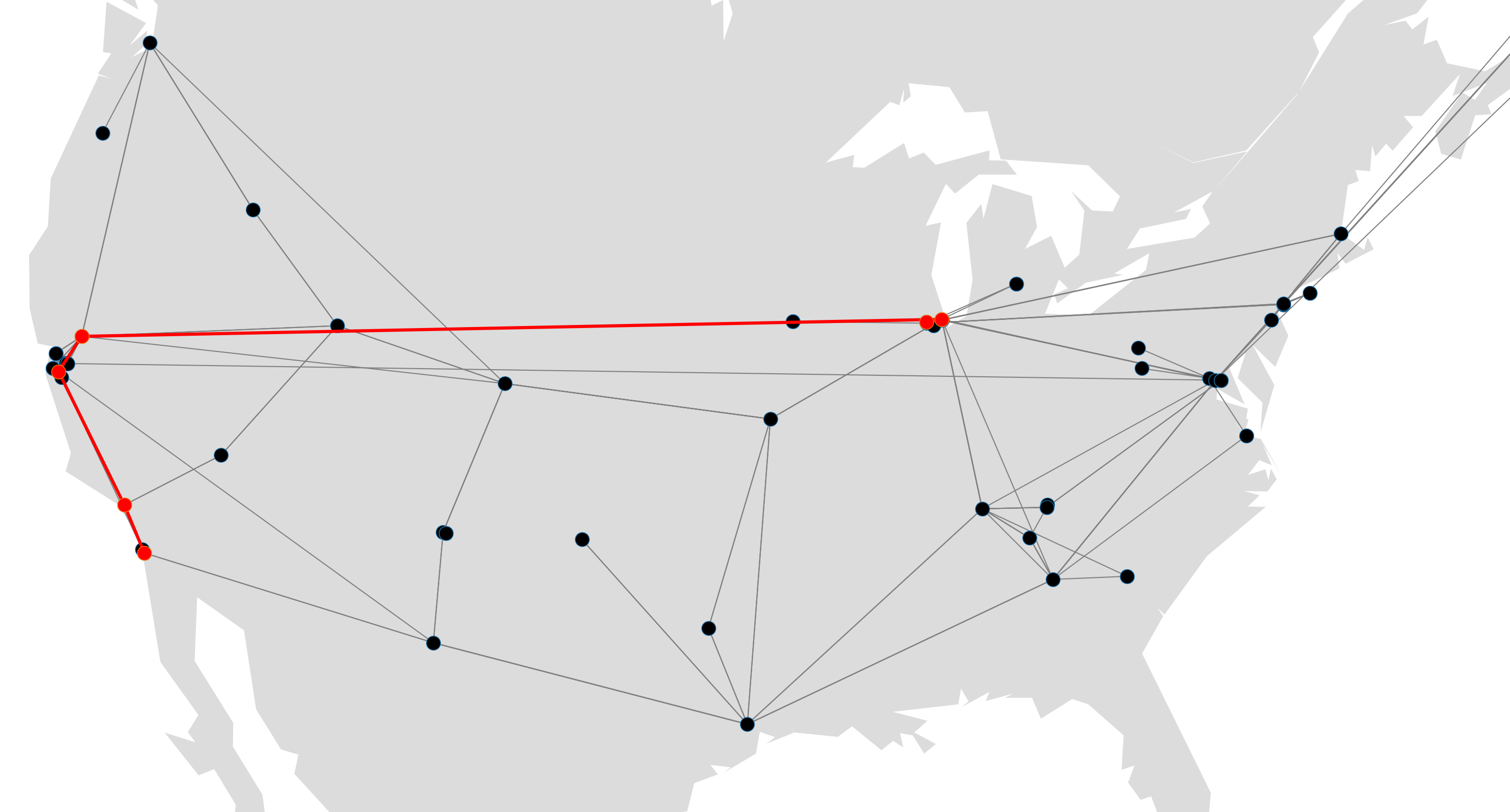}}
\caption{
  The simulated network topology of ESnet in the continental United States is plotted.
  Links are shown as grey lines and sites and routers are black dots.
  The following route between UCSD and Fermilab is plotted as a red line, and the nodes involved are highlighted in red: San Diego $\rightarrow$ Sunnyvale $\rightarrow$ Sacremento $\rightarrow$ Chicago $\rightarrow$ Fermilab. 
  The route in this preliminary work was selected using Dijkstra's algorithm~\cite{Dijkstra1959, DijkstraTextbook} where all links are weighed equally. 
  This network topology, along with measured bandwidths, will be used to simulate data movement across ESnet.
}
\label{fig:sim-topology}
\end{figure}

Figure \ref{fig:sim-topology} shows the network topology of ESnet as implemented presently in this simulation. 
This topology will be used as an idealized model of the network, based on the theoretical bandwidth between nodes in ESnet as computed by experts. 
These bandwidths can then be divided amongst imitation SENSE provisions, or equally shared amongst best-effort traffic. 
With this information, we can simulate the duration of a given set of data transfers, including how they might interfere with one another. 
We expect that this model can then be used to approximate data movement under a set of experimental policies, which we can then vary to evaluate relative performance. 
Moreover, we see this initial work as a foundation upon which we can build a simulation that more closely resembles reality. 
Detailed results from these simulations will be reported in future publications.

\section{Near-term Goals}
By Supercomputing 2022, we hope to expand the bandwidth in our Caltech-UCSD testbed from currently 10 Gb/s to up to 400 Gb/s to understand scalability of the XRootD infrastructure. 
Early in 2023, we then hope to add the CMS Tier-1 at Fermi National Laboratory (FNAL) in Chicago, and the CMS Tier-2 at University of Nebraska Lincoln (UNL) to our prototype testbed. 
We see this as a step towards contributing to the WLCG Data Challenge DC23, which is expected in either late 2023 or early 2024. 
We expect to contribute to DC23 via the testbed of four sites without integration into the production Rucio instance of CMS. 
We then hope to accomplish that integration into production by DC25.  

Long term, we expect to develop and validate new features on the testbed before we migrate them into the production instance of Rucio for CMS.

\section*{Acknowledgments}
This ongoing work is partially supported by the US National Science Foundation (NSF) Grants OAC-2030508, OAC-1841530, OAC-1836650, MPS-1148698, and PHY-1624356.
In addition, the development of SENSE is supported by the US Department of Energy (DOE) Grants DE-SC0015527, DE-SC0015528, DE-SC0016585, and FP-00002494.
Finally, this work would not be possible without the significant contributions of collaborators at ESnet, Caltech, and SDSC.

\bibliographystyle{IEEEtran}
\bibliography{conference}

\end{document}